\documentclass[showpacs,amssymb,preprint]{revtex4}
\usepackage{graphicx}% Include figure files
\usepackage{dcolumn}% Align table columns on decimal point
\usepackage{bm}% bold math
\begin{document}
\title{Utility Function from Maximum Entropy Principle}
\author{Amir H. Darooneh}
\affiliation{Department of Physics, Zanjan University, P.O.Box
45196-313, Zanjan, Iran.}
\email{darooneh@mail.znu.ac.ir}
\date{\today}
\begin{abstract}
We apply the maximum entropy principle to economic systems in
equilibrium and find the density function for the market's wealth.
This is the same as price density which is used for insurance
pricing. The risk aversion parameter of the agent then it's
utility function with respect to this density is derived.
\end{abstract}

\pacs{89.65.Gh, 05.20.-y} \maketitle

\section{Introduction and Summary}
Utility functions one of the most important and useful concepts in
the economics which show the tendency of an economic agent for
acquiring the more benefit. It is originated from the concept of
potential energy in the physics \cite{bjn}. There is no direct way
for finding the utility function of an agent with respect to it's
financial condition, namely its wealth. Different companies find
their utility function from analysis of trading data.

Recently the insurance market which is one of the important branch
of economy have attracted the attention of physicists
\cite{fd,d1,d2,d3,d4,d5}. The statistical mechanics concepts
specially the maximum entropy principle is used for pricing the
insurance \cite{d4,d5}. The well known results on economic premium
calculation \cite{b1,b2} are retrieved.

In the next section we follow the work of Darooneh \cite{d5} to
obtain the price density based on the maximum entropy principle
then in last section we apply it for multi agents model of
insurance market \cite{b1,d5}. Finally the utility function will
be derived. The main assumption here is coincidence of concept of
the equilibrium in physics and economics. This is not strange
because both approaches to equilibrium have the same results
\cite{d4,d5,b1}. However the problem of equilibrium for multi part
systems is under the investigation by author.

The power of statistical mechanics enables us to extend and apply
this method for calculation of utility function in other cases
such as finite markets and take into the account the effects of
other constraint in the market.

\section{The Maximum Entropy Method in Economics}
The risky events affect the financial market. The randomness in
the market as a consequence of the risks will be increased when
the times goes forward. The state of the market with maximum
randomness is called equilibrium. In equilibrium state we lose the
most information about the status of economic agents in the
market, in this respect the adoption of a strategy for trading
becomes more cumbersome. The important task of a trader, and also
an economist, is prediction state of the market. This is carried
out by calculation of distribution function for falling of the
market into a possible equilibrium state. The maximum entropy
principle appears as the best way when we make inference about an
unknown distribution based only on the incomplete information. The
Jaynes entropy may be written as \cite{j},
\begin{equation}\label{e1}
H[\varphi]=-\int_{\Omega}
\varphi(\omega)\ln\varphi(\omega)d\Pi(\omega).
\end{equation}
Where $\omega$ is an element of the risk's probability space
$\Omega$. The measure of the integral demonstrates the weight for
occurrence a random event (risk).

The distribution function $\varphi(\omega)$ is normalized
function.
\begin{equation}\label{e2}
\int_{\Omega} \varphi(\omega)d\Pi(\omega)=1.
\end{equation}
It is assumed that the average of the market's wealth in the
equilibrium state should be constant.
\begin{equation}\label{e3}
<W>=\int_{\Omega} \varphi(\omega)W(\omega)d\Pi(\omega)=Const.
\end{equation}
This is not surprising assumption because money exchanging between
different agents (market's constituents) doesn't alter the
market's wealth totally. That money which enter or leave the
market is controlled by trading strategies.

The eqs. \ref{e2} and \ref{e3} should be satisfied simultaneously
when we attempt to maximize the entropy \ref{e1}.
\begin{widetext}
\begin{equation}\label{e4}
\delta H[\varphi]+\lambda \delta\int_{\Omega}
\varphi(\omega)d\Pi(\omega)+\beta \delta \int_{\Omega}
\varphi(\omega)W(\omega)d\Pi(\omega)=0.
\end{equation}
\end{widetext}

The canonical distribution is the well known solution to above
equation \cite{p}.
\begin{equation}\label{e5}
\varphi(\omega)=\frac{e^{-\beta W(\omega)}}{\int_{\Omega}
e^{-\beta W(\omega)}d\Pi(\omega)}.
\end{equation}

The parameter $\beta$ has important roll in price density, it can
be calculated on basis of the method that is introduced in
previous works \cite{d2,d4}, but our intuition from similar case
in thermal physics tell us \cite{p},
\begin{equation}\label{e6}
\beta\approx\frac{1}{<W>}.
\end{equation}

In the case of insurance market the wealth of a typical agent;
insurer or insurant, is given by,
\begin{equation}\label{e7}
W_{i}(\omega)=W_{0i}-X_{i}(\omega)+Y_{i}(\omega)-\int_{\Omega}
\varphi(\omega)Y_{i}(\omega)d\Pi(\omega).
\end{equation}
The index $i$ indicates the different agents.

Each agent in the market will be incurred $X_{i}(\omega)$ if
$\omega$ is happening. He insured himself for the price
$\int_{\Omega} \varphi(\omega)Y_{i}(\omega)d\Pi(\omega)$ and
receives $Y_{i}(\omega)$ upon occurrence of this event.

As we mentioned before the money is only exchanged between agents
and doesn't alter the total money in the market. It is what is
called the clear condition.
\begin{equation}\label{e8}
\sum_{i}Y_{i}(\omega)=0.
\end{equation}
The market's wealth is sum of the individual wealth of the agents.
\begin{equation}\label{e9}
W(\omega)=\sum_{i}W_i(\omega)=W_0-Z(\omega).
\end{equation}
It is also true for initial wealth $W_0$ and aggregate risk
$Z(\omega)$,
\begin{eqnarray}\label{e10}
W_0&=&\sum_{i}W_{0i}\nonumber \\
Z(\omega)&=&\sum_iX_i(\omega).
\end{eqnarray}

With the aid of eq. \ref{e9} we can rewrite eq. \ref{e6} to
retrieve the B\"{u}hlmann result on economic premium calculation
\cite{b1,b2}.
\begin{equation}\label{e11}
\varphi(\omega)=\frac{e^{\beta Z(\omega)}}{\int_{\Omega} e^{\beta
Z(\omega)}d\Pi(\omega)}.
\end{equation}

In the next section we try to obtain the utility function from the
results that have been fund in this section.

\section{Utility Function}

The utility function demonstrates what an agent is interested for
making specified amounts of profit in a competitive market. The
utility function should depend on financial status of an economic
agent which is frequently described by its wealth, $u(W)$. It is
assumed that the utility function has positive first derivative,
$u'(W)>0$, to guarantee that the agent is willing the profit and
negative second derivative, $u''(W)<0$, to restrict it's avarice.
The risk aversion parameter, $\beta(W)=-u''(W)/u'(W)$, is also
involved in the utility function to scale the agent's will in the
market with respect to it's wealth.

The market reach at equilibrium upon the agents are satisfied from
their trade. In this respect the agent's utility function should
be the most in the equilibrium state. Because of the risks in the
market which induce the randomness, the equilibrium condition may
be expressed as an average form.
\begin{equation}\label{e12}
\int_{\Omega} u_{i}(W_{i}(\omega))d\Pi(\omega)=max.
\end{equation}

The only function that can be changed upon our request is
$Y_i(\omega)$ hence maximizing of the utility function means,
\begin{equation}\label{e13}
\frac{\delta}{\delta Y_i(\omega)} \int_{\Omega}
u_{i}(W_{i}(\omega))d\Pi(\omega)=0.
\end{equation}

It is not so hard to derive the following result by using
variational technique.
\begin{eqnarray}\label{e14}
u'_i(W_i(\omega))&=&\varphi(\omega)\int u'_i(W_i(\upsilon))d\Pi(\upsilon) \nonumber \\
u'_i(W_i(\omega))&=&C_i\varphi(\omega).
\end{eqnarray}
Where the derivative are respect to argument of the functions.

We have freedom to rescale the risk exchange function
$Y_i(\omega)$, up to a constant number since the agent's wealth
remains unaltered.
\begin{equation}\label{e15}
Y_i^{New}(\omega)=Y_i^{Old}(\omega)-\int_{\Omega}
\varphi(\omega)Y_{i}^{Old}(\omega)d\Pi(\omega)
\end{equation}

The risk function is also renewed to absorb the risk exchange
function.
\begin{equation}\label{e15r}
X_i^{New}(\omega)=X_i^{Old}(\omega)-Y_i^{New}(\omega).
\end{equation}

Often a loss event is incurred the market totally and the agent's
risk function depends on $\omega$ through $\zeta=Z(\omega)$ hence
it is suitable and very often necessary to deal with $\zeta$
instead of $\omega$ \cite{b2}.
\begin{eqnarray}\label{e16}
\sum X_i(\zeta)&=&\zeta \nonumber \\
\varphi(\zeta)&\sim& e^{\beta \zeta}.
\end{eqnarray}

Taking the logarithmic derivative of eq. \ref{e14} we obtain,
\begin{eqnarray}\label{e17}
-\frac{u''_i(W_i(\zeta))}{u'_i(W_i(\zeta))}X'_i(\zeta)=\frac{\varphi
'(\zeta)}{\varphi(\zeta)}.
\end{eqnarray}
From definition of the risk aversion parameter we have,
\begin{eqnarray}\label{e18}
\beta_i=\frac{\beta}{X'_i(\zeta)}.
\end{eqnarray}

If the risk function for an agent be specified then the utility
function is given by the following relation.
\begin{eqnarray}\label{e19}
u_i(\xi)=\int_0^\xi e^{-\int_0^\eta\beta_i d\zeta}d\eta.
\end{eqnarray}
In the above equation we standardize the utility function by
assuming $u_i(0)=0$ and $u'_i(0)=1$. For constant risk aversion
parameter the exponential form for utility function is obtained.

\begin{acknowledgments}
The author acknowledge Dr. Saeed for reading the manuscript and
his valuable comments. This work has been supported by the Zanjan
university research program on Non-Life Insurance Pricing No.
8243.
\end{acknowledgments}

\bibliographystyle{}

\end{document}